\documentclass[aps,preprint,groupedaddress]{revtex4}

\usepackage{graphics,epsfig}

\begin{document}

\title{Bounds for FCNC  and the Pseudoscalar Higgs mass in the General
Two Higgs Doublet  Model type III using $g-2$ muon factor}
\author{Rodolfo A. Diaz}
\email[radiaz@ciencias.ciencias.unal.edu.co]{}
\author{R. Martinez}
\email[romart@ciencias.ciencias.unal.edu.co]{}                        
\author{J-Alexis Rodriguez} 
\email[alexro@ciencias.ciencias.unal.edu.co]{}
%\homepage[]{Your web page}
%\thanks{}
%\altaffiliation{}
\affiliation{Departamento de Fisica, Universidad Nacional de Colombia\\
Bogota, Colombia}

\begin{abstract}
Current muon anomalous magnetic moment $a_{\mu }\;$ data have challenged
Standard Model (SM) and seem to open a window for new physics. Since the difference
between SM and experimental predictions is approximately $2.6\sigma $. In
the framework of the General Two Higgs Doublet Model (2HDM), we calculate
the muon anomalous magnetic moment to get lower and upper bounds for the
Flavour Changing (FC) Yukawa couplings in the leptonic sector. We also
obtain lower bounds for the mass of the Pseudoscalar Higgs ($m_{A^{0}}$) as
a function of the parameters of the model.
\end{abstract}

\maketitle

Current muon anomalous magnetic moment $a_{\mu }\;$ data have challenged
Standard Model (SM) and seem to open a window for new physics. Due to the

high precision in $a_{\mu }$ value, it gives very restrictive bounds on
physics beyond the SM. Although $a_{e}$ measurement is about 350 more
precise \cite{marciano}, $a_{\mu }\;$is much more sensitive to New Physics
since contributions to $a_{l}\;$are usually proportional to $m_{l}^{2}$.

The most accurate measurement of $a_{\mu }\;$hitherto, has been provided by
the Brookhaven Alternating Gradient Syncrotron \cite{g2 positive}. Their
data have an error one third that of the combined previous data \cite{Caso},
ref \cite{g2 positive} reports 
\begin{equation}
a_{\mu ^{+}}=11659202\left( 14\right) \left( 6\right) \times 10^{-10}.
\label{g2 muon exp}
\end{equation}
On the other hand, SM predictions for $a_{\mu }$ has been estimated taking
into account the contributions from QED, Hadronic loops and electroweak
corrections. The final current result is \cite{marciano, g2 positive} 
\begin{equation}
a_{\mu }^{SM}=11659159.7\left( 6.7\right) \times 10^{-10}.
\label{g2 muon SM}
\end{equation}

Taking into account (\ref{g2 muon SM}) is obtained 
\begin{equation}
\Delta a_{\mu }^{NP}=a_{\mu }^{\exp }-a_{\mu }^{SM}=42.6\left( 16.5\right)
\times 10^{-10},
\end{equation}
where $a_{\mu }^{\exp }\;$is the world average experimental value \cite
{marciano}. Consequently at $90\%$ C.L. 
\begin{equation}
21.5\times 10^{-10}\leq \Delta a_{\mu }^{NP}\leq 63.7\times 10^{-10}.
\label{Room for NP}
\end{equation}
$\Delta a_{\mu }^{NP}\;$gives the room available for New Physics, so $a_{\mu
}^{exp}$ differs from $a_{\mu }^{SM}$ approximately in $2.6\sigma $.
Therefore, physics beyond the SM is needed to achieve an acceptable
theoretical experimental agreement. The most studied contributions to $%
a_{\mu }\;$has been carried out in the framework of radiative muon mass
models as well as the Minimal Supersymmetric Standard Model (MSSM), E$_{6}\;$%
string-inspired models, and extensions of MSSM with an extra singlet \cite
{estatus}.

Moreover, a very interesting suggestion to conciliate the new experimental
data with theoretical predictions is to consider models that includes FCNC
at tree level. Interactions involving FCNC are forbidden at tree level in
the SM, but could be present at one loop level as in the case of $%
b\rightarrow s\gamma $ \cite{bsg}, $K^{0}\rightarrow \mu ^{+}\mu ^{-}$ \cite
{kmm}, $K^{0}-\overline{K}^{0}$ \cite{koko}, $t\rightarrow c\gamma $ \cite
{tcg} etc. Many extensions of the SM permit FCNC at tree level.\ For
example, the introduction of new representations of fermions different from
doublets produce them by means of the Z-coupling \cite{2}. Additionally,
they are generated at tree level by adding a second doublet to the SM \cite
{wolf}, such couplings can be gotten as well in SUSY theories without
R-parity. Some other important new sources for FCNC might be provided by a
muon collider, as the processes $\mu \mu \rightarrow \mu \tau (e\tau )\;$
mediated by Higgs exchange \cite{Workshop}, \cite{SherCollider}, which
produce Lepton Flavor Violation (LFV).

However, there are several mechanisms to avoid FCNC at tree level. Glashow
and Weinberg \cite{gw} proposed a discrete symmetry to supress them in the
Two Higgs Doublet Model (2HDM) which is the simplest one that exhibits these
rare processes at tree level. There are two kinds of models which are
phenomenologically plausible with the discrete symmetry imposed. In the
model type I, one Higgs Doublet provides masses to the up-type and down-type
quarks, simultaneously. In the model type II, one Higgs doublet gives masses
to the up-type quarks and the other one to the down-type quarks. But the
discrete symmetry \cite{gw} is not compulsory and both doublets may generate
the masses of the quarks of up-type and down-type simultaneously, in such
case we are in the model type III \cite{III}. It has been used to search for
FCNC at tree level \cite{ARS}, \cite{Sher91}.

Recently, the 2HDM type III has been discussed and classified \cite{us},
depending on how the basis for the vacuum expectation values (VEV) are
chosen and according to the way in which the flavor mixing matrices are
rotated. In brief, the reference \cite{us} shows that there are two types of
rotations which generate four different lagrangians in the quark sector and
two different ones in the leptonic sector. The well known 2HDM types I and
II, could be generated from them in the limit in which the FC vertices
vanish. It has been pointed out that the phenomenology of the 2HDM type III
is highly sensitive to the rotation used for the mixing matrices.

In this paper, we calculate the contributions to $\Delta a_{\mu }^{NP}$
coming from the 2HDM, which includes FCNC at tree level. We will constrain
the FC vertex involving the second and third charged leptonic sector by
using the result for $\Delta a_{\mu }^{NP}$, equation (\ref{Room for NP}).
Additionally, we get lower bounds on the Pseudoscalar Higgs mass by taking
into account the lower experimental value of $\Delta a_{\mu }^{NP}\;$ at $%
90\%$ CL and making reasonable assumptions on the FC vertex.

The Yukawa's Lagrangian for the 2HDM type III, is as follow 
\begin{eqnarray}
-\pounds _{Y} &=&\eta _{ij}^{U}\overline{Q}_{iL}\widetilde{\Phi }%
_{1}U_{jR}+\eta _{ij}^{D}\overline{Q}_{iL}\Phi _{1}D_{jR}+\eta _{ij}^{E}%
\overline{l}_{iL}\Phi _{1}E_{jR}  \nonumber \\
&+&\xi _{ij}^{U}\overline{Q}_{iL}\widetilde{\Phi }_{2}U_{jR}+\xi _{ij}^{D}%
\overline{Q}_{iL}\Phi _{2}D_{jR}+\xi _{ij}^{E}\overline{l}_{iL}\Phi
_{2}E_{jR}+h.c.  \label{Yukawa}
\end{eqnarray}
where $\Phi _{1,2}\;$are the Higgs doublets,$\;\eta _{ij}\;$and $\xi _{ij}\;$%
are non-diagonal $3\times 3\;$matrices and $i$, $j$ are family indices. In
this work, we are interested only in neutral currents in the leptonic
sector. We consider a CP-conserving model in which both Higgs doublets
acquire a VEV, 
\begin{equation}
\left\langle \Phi _{1}\right\rangle _{0}=\left( 
\begin{array}{c}
0 \\ 
v_{1}/\sqrt{2}
\end{array}
\right) \;\;,\;\;\left\langle \Phi _{2}\right\rangle _{0}=\left( 
\begin{array}{c}
0 \\ 
v_{2}/\sqrt{2}
\end{array}
\right) .  \nonumber
\end{equation}
The neutral mass eigenstates are given by \cite{moda} 
\begin{eqnarray}
\left( 
\begin{array}{c}
G_{Z}^{0} \\ 
A^{0}
\end{array}
\right) &=&\left( 
\begin{array}{cc}
\cos \beta & \sin \beta \\ 
-\sin \beta & \cos \beta
\end{array}
\right) \left( 
\begin{array}{c}
\sqrt{2}Im\phi _{1}^{0} \\ 
\sqrt{2}Im\phi _{2}^{0}
\end{array}
\right) ,  \nonumber \\
\left( 
\begin{array}{c}
H^{0} \\ 
h^{0}
\end{array}
\right) &=&\left( 
\begin{array}{cc}
\cos \alpha & \sin \alpha \\ 
-\sin \alpha & \cos \alpha
\end{array}
\right) \left( 
\begin{array}{c}
\sqrt{2}Re\phi _{1}^{0}-v_{1} \\ 
\sqrt{2}Re\phi _{2}^{0}-v_{2}
\end{array}
\right)  \label{Autoestados masa Higgs}
\end{eqnarray}
where $\tan \beta =v_{2}/v_{1}\;$and $\alpha \;$is the mixing angle of the
CP-even neutral Higgs sector. $G_{Z}\;$is the would-be Goldstone boson of $Z$
and $A^{0}\;$is the CP-odd neutral Higgs.

Now, to convert the Lagrangian (\ref{Yukawa}) into mass eigenstates we make
the unitary transformations 
\begin{equation}
E_{L,R}=\left( V_{L,R}\right) E_{L,R}^{0}\;\;  \label{Down transf}
\end{equation}
from which we obtain the mass matrix 
\begin{equation}
M_{E}^{diag}=V_{L}\left[ \frac{v_{1}}{\sqrt{2}}\eta ^{E,0}+\frac{v_{2}}{%
\sqrt{2}}\xi ^{E,0}\right] V_{R}^{\dagger }\;\;,  \label{Masa down}
\end{equation}
where $M_{E}^{diag}$ is the diagonal mass matrix for the three lepton
families. From (\ref{Masa down}) we can solve for $\xi ^{E,0}\;$obtaining 
\begin{equation}
\xi ^{E,0}=\frac{\sqrt{2}}{v_{2}}V_{L}^{\dagger }M_{E}^{diag}V_{R}-\frac{%
v_{1}}{v_{2}}\eta ^{E,0}  \label{Rotation Id}
\end{equation}
which we call a rotation of type I. Replacing it into (\ref{Yukawa}), the
expanded Lagrangian for the neutral leptonic sector is 
\begin{eqnarray}
-\pounds _{Y\left( E\right) }^{\left( I\right) } &=&\frac{g}{2M_{W}\sin
\beta }\overline{E}M_{E}^{diag}E\left( \sin \alpha H^{0}+\cos \alpha
h^{0}\right)  \nonumber \\
&+&\frac{ig}{2M_{W}}\overline{E}M_{E}^{diag}\gamma _{5}EG^{0}+\frac{ig\cot
\beta }{2M_{W}}\overline{E}M_{E}^{diag}\gamma _{5}EA^{0}  \nonumber \\
&-&\frac{1}{\sqrt{2}\sin \beta }\overline{E}\eta ^{E}E\left[ \sin \left(
\alpha -\beta \right) H^{0}+\cos \left( \alpha -\beta \right) h^{0}\right] 
\nonumber \\
&-&\frac{i}{\sqrt{2}\sin \beta }\overline{E}\eta ^{E}\gamma _{5}EA^{0}+h.c.
\label{Yukawa 1ad}
\end{eqnarray}
where the superindex $(I)\;$refers to the rotation type I. It is easy to
check that Lagrangian (\ref{Yukawa 1ad}) is just the one in the 2HDM type I\ 
\cite{moda}, plus some FC interactions. Therefore, we obtain the Lagrangian
of the 2HDM type I from eq (\ref{Yukawa 1ad}) by setting $\eta ^{E}=0.\;$ In
this case it is clear that when $\tan \beta \rightarrow 0$ then $\eta ^{E}$
should go to zero, in order to have a finite contribution for FCNC at tree
level.

On the other hand, from (\ref{Masa down}) we can also solve for $\eta
^{E,0}\;$instead of $\xi ^{E,0}$, to get 
\begin{equation}
\eta ^{E,0}=\frac{\sqrt{2}}{v_{1}}V_{L}^{\dagger }M_{E}^{diag}V_{R}-\frac{%
v_{2}}{v_{1}}\xi ^{E,0}  \label{Rotation IId}
\end{equation}
which we call a rotation of type II. Replacing it into (\ref{Yukawa}) the
expanded Lagrangian for the neutral leptonic sector is 
\begin{eqnarray}
-\pounds _{Y(E)}^{(II)} &=&\frac{g}{2M_{W}\cos \beta }\overline{E}%
M_{E}^{diag}E\left( \cos \alpha H^{0}-\sin \alpha h^{0}\right)  \nonumber \\
&+&\frac{ig}{2M_{W}}\overline{E}M_{E}^{diag}\gamma _{5}EG^{0}-\frac{ig\tan
\beta }{2M_{W}}\overline{E}M_{E}^{diag}\gamma _{5}EA^{0}  \nonumber \\
&+&\frac{1}{\sqrt{2}\cos \beta }\overline{E}\xi ^{E}E\left[ \sin \left(
\alpha -\beta \right) H^{0}+\cos \left( \alpha -\beta \right) h^{0}\right] 
\nonumber \\
&+&\frac{i}{\sqrt{2}\cos \beta }\overline{E}\xi ^{E}\gamma _{5}EA^{0}+h.c.
\label{Yukawa 2ad}
\end{eqnarray}
The Lagrangian (\ref{Yukawa 2ad}) coincides with the one of the 2HDM type II 
\cite{moda}, plus some FC interactions. So, the Lagrangian of the 2HDM type
II is obtained setting $\xi ^{E}=0.\;$ In this case it is clear that when $%
\tan \beta \rightarrow \infty $ then $\xi ^{E}$ should go to zero, in order
to have a finite contribution for FCNC at tree level.

In the present report, we calculate $\Delta a_{\mu }^{NP}$ in the 2HDM with
FC interactions. If we neglect the muon mass, the contribution at one loop
from all Higgses is given by 
\begin{equation}
\Delta a_{\mu }^{NP}=\frac{m_{\mu }m_{l}}{16\pi ^{2}}\sum_{i}F\left(
m_{H_{i}},m_{l}\right) a_{i}^{2}\;,  \label{Anomal}
\end{equation}
where 
\begin{equation}
F\left( m_{H_{i}},m_{l}\right) =\allowbreak \frac{3+\widehat{m}%
_{H_{i}}^{2}\left( \widehat{m}_{H_{i}}^{2}-4\right) +2\ln \widehat{m}%
_{H_{i}}^{2}}{m_{H_{i}}^{2}\left( 1-\widehat{m}_{H_{i}}^{2}\right) ^{3}}
\label{anomalo}
\end{equation}
$\allowbreak $ \allowbreak with $\widehat{m}_{H_{i}}=m_{l}/m_{H_{i}}$\ and $%
m_{l}\;$is the mass of the lepton running into the loop. The sum is over the
index $i=m_{h^{0}},m_{H^{0}},m_{A^{0}}$. The coefficients $a_{i}$ are the
Feynman rules for the FC couplings involved.

If we take into account the experimental data (\ref{Room for NP}), we get
some lower and upper bounds on the mixing vertex $\eta \left( \xi \right)
_{\mu \tau }\;$for the rotations of type I (II). In figure 1, we display
lower and upper bounds for the FC vertices as a function of $\tan \beta $
for both types of rotations with $m_{h^{0}}=m_{H^{0}}=150$ GeV and $%
m_{A^{0}}\rightarrow \infty $. In the first case, rotation type I, the
allowed region for $\eta _{\mu \tau }$ is $0.07\leq \eta _{\mu \tau }\leq
0.13$ for large values of $\tan \beta $. Meanwhile, for rotation type II,
the allowed region for small $\tan \beta $ is the same. From Lagrangian
(11), which describes rotation type I, we can see that when $\tan \beta
\rightarrow 0$, $\eta _{\mu \tau }$ should go to zero as well to mantain a
finite contribution to $\Delta a_{\mu }$. This behaviour can be seen from
figure 1. For rotation type II occurs the same but in the limit $\tan \beta
\rightarrow \infty $.

In figure 2, we show lower and upper bounds for the FC vertex as a function
of $m_{H^{0}}$ for rotation of type II when $m_{h^{0}}=m_{H^{0}}$ and $%
m_{A^{0}}\rightarrow \infty $. We see that the larger value for $\tan \beta $
the smaller value of $\xi _{\mu \tau }$ . We only consider the case of
rotation type II because there is a complementary behaviour between both
rotations as could be seen in figure 1. In particular, for $\tan \beta =1,$
the behaviour of the bounds for both rotations is the same.

Observe that according to the Feynman rules from (\ref{Yukawa 1ad}) and (\ref
{Yukawa 2ad}), the Scalar (Pseudoscalar) contribution to $\Delta a_{\mu
}^{NP}\;$eq. (\ref{Anomal},\ref{anomalo}) is positive (negative). Such fact
permits us to impose lower bounds on the Pseudoscalar Higgs mass, by using
the lower limit in eq. (\ref{Room for NP}). According to this equation the
room for new physics from $g-2$ muon factor is positive definite, and it is
a new feature from most updated results \cite{g2 positive}.

Now, to take into account the experimental value (\ref{Room for NP}), we
should make a supposition about the value of the FC vertex. A reasonable
assumption consists of taking the geometric average of the Yukawa couplings 
\cite{Cheng Sher} i.e. $\eta \left( \xi \right) _{\mu \tau }\approx
2.5\times 10^{-3}$. Additionally, we shall use also the values $\eta \left(
\xi \right) _{\mu \tau }\approx 2.5\times 10^{-2}$ and$\;\eta \left( \xi
\right) _{\mu \tau }\approx 2.5\times 10^{-4}\;$which are one order of
magnitude larger and smaller than the former. Using these suppositions and
the experimental value (\ref{Room for NP}) we get lower bounds for $%
m_{A^{0}}\;$ and they are plotted in figures (3)-(5).

Figure 3 displays $m_{A^{0}}\;$vs$\;\tan \beta \;$using rotation type II
with the three values of $\xi _{\mu \tau }\;$mentioned above and setting $%
m_{h^{0}}=m_{H^{0}}\;$with $m_{h^0}=110,\;300$ GeV. It could be seen that in
the limit of large $\tan \beta ,\;$the lower limit reduces to$%
\;m_{A^{0}}\approx m_{h^0}$. The same behavior can be seen in rotation type
I but the bound $m_{A^{0}}\approx m_{h^0}$ is gotten in the limit of small $%
\tan \beta .$ We also see that the smaller value of $\xi_{\mu \tau }\;$the
stronger lower limit for $m_{A^{0}}$.

Figure 4 shows $m_{A^{0}}\;$vs$\;m_{H^{0}}\;$with $\xi _{\mu \tau
}=2.5\times 10^{-3},2.5\times 10^{-2}\;$and $\tan \beta =1$, using $%
m_{h^{0}}=m_{H^{0}}\;$and$\;m_{h^{0}}=110$ GeV. With this settings, the
value $\xi _{\mu \tau }=2.5\times 10^{-4}\;$ is excluded. Using such
specific arrangements, the bounds are identical in both types of rotations.

In figure 5 we suppose that $m_{h^{0}}=110$ GeV,$\;m_{H^{0}}=300$ GeV. The
figure above shows the sensitivity of lower bounds on $m_{A^{0}}\;$ with the
mixing angle $\alpha ,\;$for rotation type II, taking $\tan \beta =1$. The
value $\xi _{\mu \tau }=2.5\times 10^{-4}\;$is excluded again. The
constraints are very sensitive \ to the $\alpha \;$mixing angle for $\xi
_{\mu \tau }=2.5\times 10^{-3}\;$ but rather insensitive for $\xi _{\mu \tau
}=2.5\times 10^{-2}$. The figure below shows $m_{A^{0}}\;$vs\ $\tan \beta \;$%
for $m_{h^{0}}=110$ GeV, $m_{H^{0}}=300$ GeV, $\alpha =\pi /6,\;$ for
rotation type II and considering the same three values of $\xi _{\mu \tau
}.\;$The $m_{A^{0}}$ lower asymptotic limit for large $\tan \beta \;$is
approximately $m_{h^{0}}$.

In conclusion, we have found lower and upper bounds for the FC vertex $\eta
\left( \xi \right) _{\mu \tau }\;$in the context of the general 2HDM by
using the allowed range for $\Delta a_{\mu }^{NP}\;$ at $90\%$ CL and$\;$%
utilizing several sets of values for the parameters of the model.
Additionally, in the limit $m_{A^{0}}\rightarrow \infty $, we get that for
small (large) values of $\tan \beta \;$the allowed range for the FC vertex $%
\eta _{\mu \tau }\left( \xi _{\mu \tau }\right) \;$becomes narrower, and
both upper and lower bounds go to zero in the rotation of type I (II).

On the other hand, we have gotten lower bounds on the Pseudoscalar Higgs
mass of the 2HDM coming from the $g-2\;$muon factor, by using the
experimental value of $\Delta a_{\mu }^{NP}\;$and making reasonable
assumptions on the FC vertex $\eta \left( \xi \right) _{\mu \tau }$.
Specifically, we have taken for $\eta \left( \xi \right) _{\mu \tau }$ the
geometric average of the Yukawa couplings, and we also utilized values one
order of magnitude larger and one order of magnitude smaller. Taking these
three values for the FC vertex we find that the smaller value for $\eta
\left( \xi \right) _{\mu \tau }\;$the more stringent lower bounds for $%
m_{A^{0}}\;$. Additionally, assuming $m_{H^{0}}=m_{h^{0}},\;$we show that in
the limit of small (large) $\tan \beta \;$the lower bound of $m_{A^{0}}\;$%
becomes merely $m_{A^{0}}\approx m_{h^{0}}\;$for rotation of type I (II). In
the case of different scalar masses, there is still a lower asymptotic limit
for $m_{A^{0}}$. Notwithstanding, these lower constraints on $m_{A^{0}}\;$%
should be consider carefully, since for $\eta \left( \xi \right) _{\mu \tau
}\;$we can only make reasonable estimations but they are unknown so far.

This work was supported by COLCIENCIAS, DIB and DINAIN.

\begin{figure}[h]
\begin{center}
\includegraphics[angle=0, width=10cm]{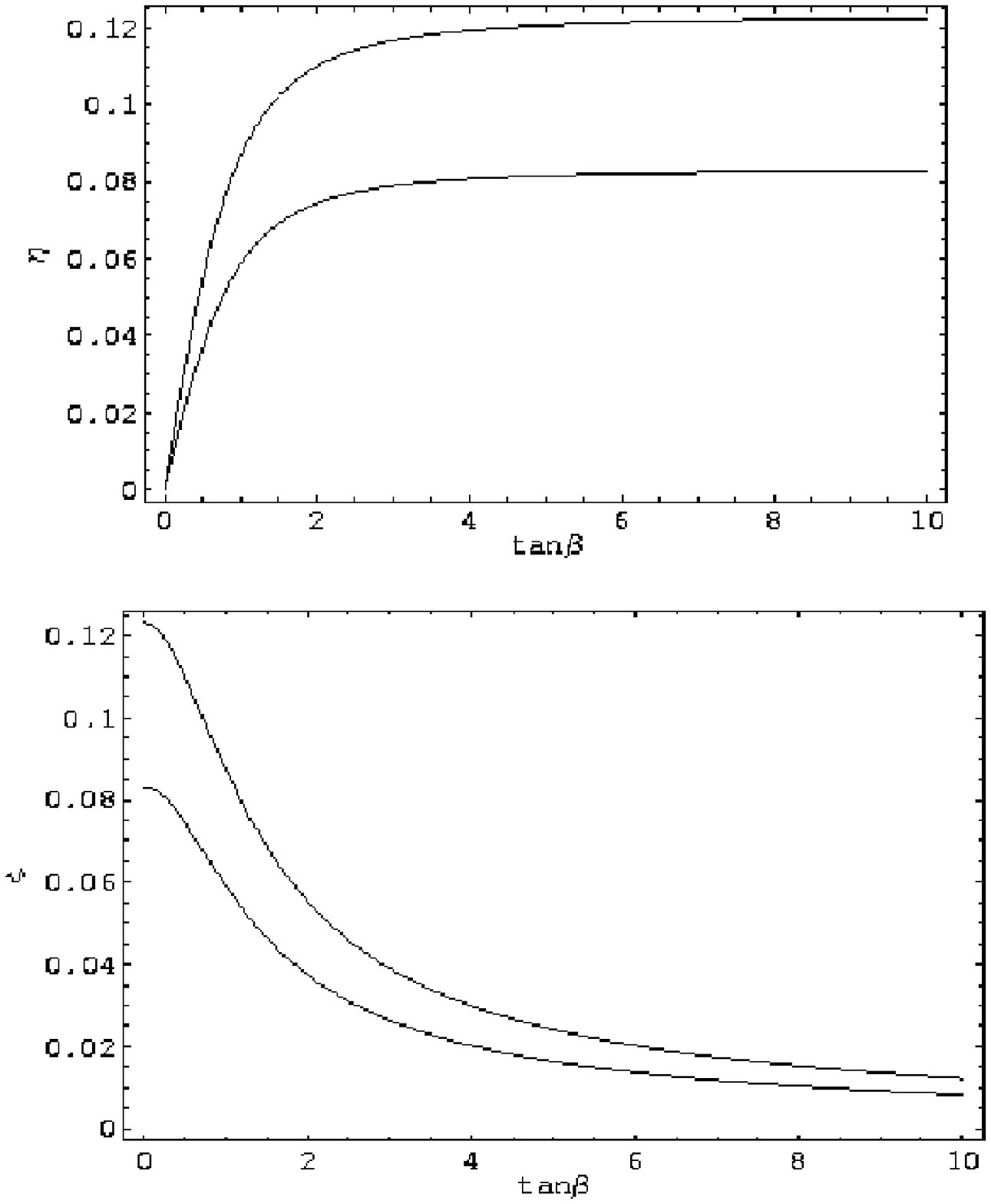}
\end{center}
\caption{ Lower and upper bounds for $\protect\eta _{\protect\mu \protect\tau
}\left( \protect\xi _{\protect\mu \protect\tau}\right) \;$vs\ tan$\protect%
\beta ,\;$for rotations I and II using $m_{h^{0}}=m_{H^{0}}=150$ GeV and $%
m_{A^{0}}\rightarrow \infty .\;$}
\label{Fig. 1}
\end{figure}

\begin{figure}[h]
\begin{center}
\includegraphics[angle=0, width=10cm]{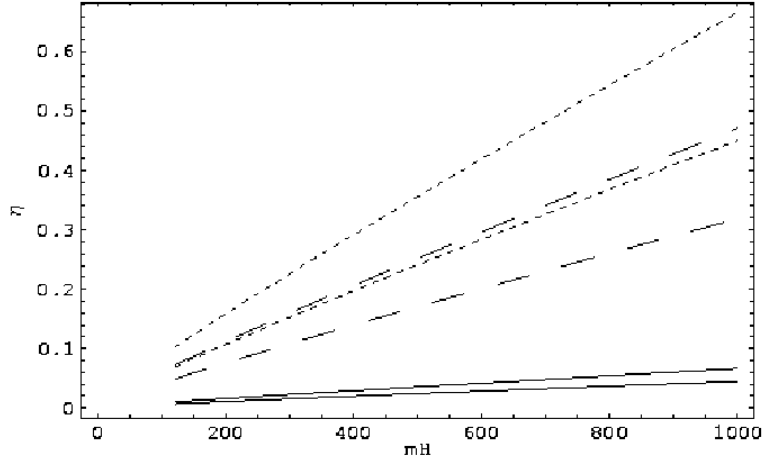}
\end{center}
\caption{ Lower and upper bounds for $\protect\xi_{\protect\mu \protect\tau
} $ vs \ $m_{H^{0}},\;$ for rotation of type II, taking$%
\;m_{h^{0}}=m_{H^{0}}\; $and $m_{A^{0}}\rightarrow \infty $, the pair of
short dashed lines correspond to $\tan \protect\beta =0.1,\;$the long dashed
lines are for $\tan \protect\beta =1$,\ and the solid lines are for $\tan 
\protect\beta =30. $}
\label{Fig. 2}
\end{figure}

\begin{figure}[h]
\begin{center}
\includegraphics[angle=0, width=10cm]{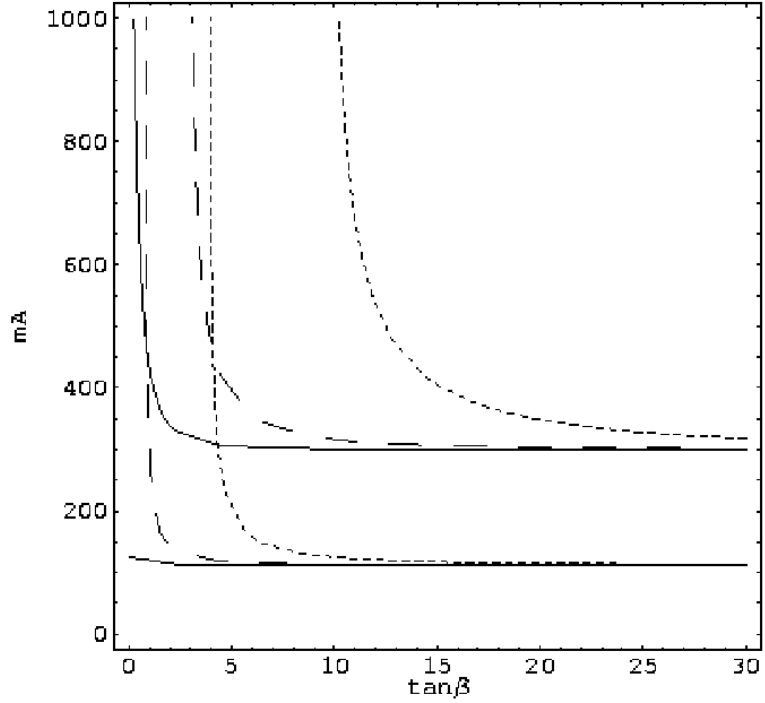}
\end{center}
\caption{ Contour plot of $m_{A^{0}}\;$vs$\;\tan \protect\beta \;$using
rotation type II and assuming $m_{h^{0}}=m_{H^{0}}$. Short dashed lines
correspond to $\protect\xi _{\protect\mu \protect\tau }=2.5\times 10^{-4}\;$%
for $m_{H^0}=110\;$ GeV (below) and $m_{H^0}=300\;$ GeV (above). Long dashed
lines correspond to $\protect\xi _{\protect\mu \protect\tau }=2.5\times
10^{-3}\;$for $m_{H^0}=110\; $ GeV (below) and $m_{H^0}=300\;$ GeV (above).
Finally, solid lines correspond to $\protect\xi _{\protect\mu \protect\tau
}=2.5\times 10^{-2}\;$for $m_{H^0}=110\;$ GeV (below) and $m_{H^0}=300\;$
GeV (above).}
\label{Fig. 3}
\end{figure}

\begin{figure}[h]
\begin{center}
\includegraphics[angle=0, width=10cm]{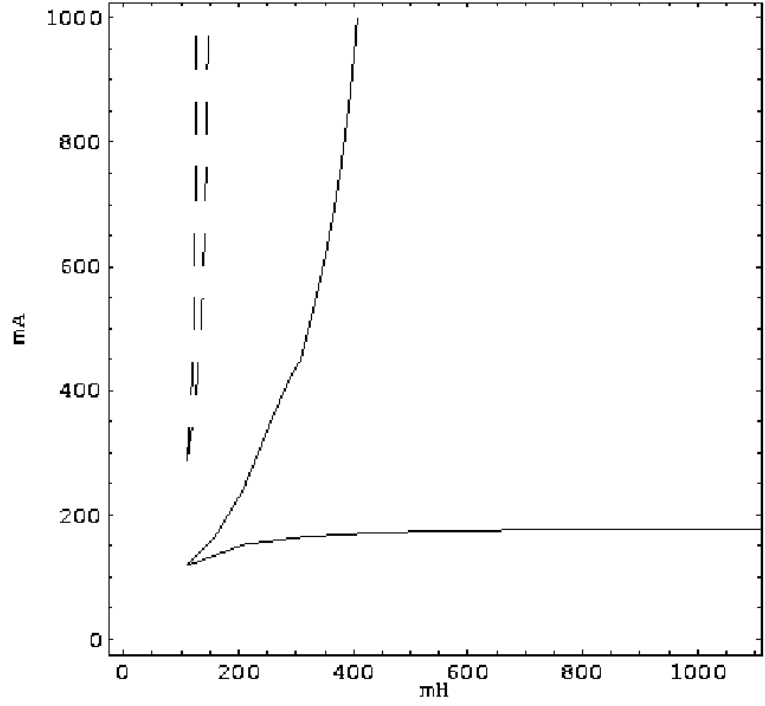}
\end{center}
\caption{ Contour plot of$\;m_{A^{0}}\;$vs$\;m_{h^{0}}\;$setting $\tan 
\protect\beta =1,\;$Long dashed lines correspond to $\protect\xi _{\protect%
\mu \protect\tau }=2.5\times 10^{-3}\;$for $m_{h^{0}}=110\;$ GeV (rigth) and 
$m_{h^{0}}=m_{H^{0}}\; $(left). Solid lines correspond to $\protect\xi _{%
\protect\mu \protect\tau }=2.5\times 10^{-2}\;$for $m_{h^{0}}=110\;$ GeV
(below) and $m_{h^{0}}=m_{H^{0}}\;$(above).}
\label{Fig. 4}
\end{figure}

\begin{figure}[h]
\begin{center}
\includegraphics[angle=0, width=8cm]{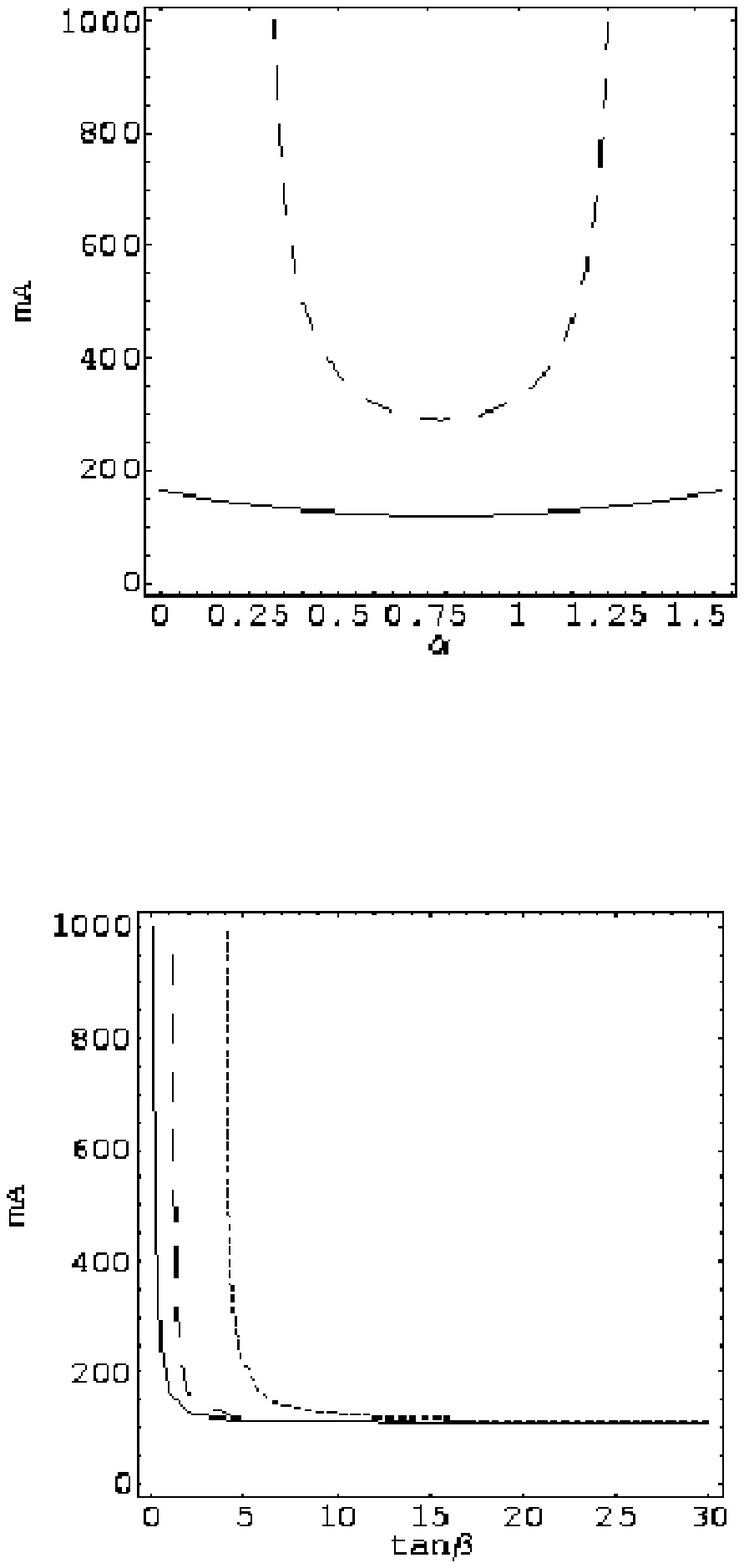}
\end{center}
\caption{(top) Contour plot of $m_{A^{0}}\;$vs\ $\protect\alpha ,\;$for
rotation type II, with $m_{h^{0}}=110$ GeV,$\;m_{H^{0}}=300$ GeV and $\tan 
\protect\beta =1$. Dashed line correspond to $\protect\xi _{\protect\mu 
\protect\tau }=2.5\times 10^{-3}$ ,\ solid line correspond to $\protect\xi _{%
\protect\mu \protect\tau }=2.5\times 10^{-2}$. (bottom) Contour plot of $%
m_{A^{0}}\;$vs\ $\tan \protect\beta \;$for $m_{h^{0}}=110$ GeV, $%
m_{H^{0}}=300$ GeV, $\protect\alpha =\protect\pi /6,\;$and for rotation type
II. Short dashed line correspond to $\protect\xi _{\protect\mu \protect\tau
}=2.5\times 10^{-4}$, long dashed line correspond to $\protect\xi _{\protect%
\mu \protect\tau }=2.5\times 10^{-3}$, and solid line correspond to $\protect%
\xi _{\protect\mu \protect\tau }=2.5\times 10^{-2}.\;$}
\label{Fig. 5}
\end{figure}

\end{document}